\documentclass[aps, twocolumn, pra]{revtex4}
\usepackage{graphicx}
\usepackage{epstopdf}

\renewcommand{\phi}{\varphi}
\newcommand{\eps}{\varepsilon}
\newcommand{\Cl}{\chi_\subs{line}}
\newcommand{\Ch}{\chi_\subs{hom}}

\newcommand{\Che}{\chi_\subs{het}}

\newcommand{\Ct}{\chi_\subs{tot}}

\newcommand{\Chi}{\chi_{\subs{h}}}
\newcommand{\subs}[1]{{\mbox{\tiny #1}}}
\newcommand{\Velec}{\ensuremath{v_\subs{el}}}
\newcommand{\IAB}{I_{\subs{AB}}}
\newcommand{\IBE}{I_{\subs{BE}}}
\newcommand{\xBE}{\chi_{\subs{BE}}}
\newcommand{\VA}{V_{\subs{A}}}
\newcommand{\VB}{V_{\subs{B}}}
\begin{document}

\title[Optical preamplifiers for improving CVQKD systems]{Improvement of continuous-variable quantum key distribution systems by using optical preamplifiers}

\author{S Fossier$^{1,2}$, E Diamanti$^2$, T Debuisschert$^1$, R Tualle-Brouri$^2$ and P Grangier$^2$}
\address{$^1$Thales Research \& Technology France, RD 128, 91767 Palaiseau Cedex, France}
\address{$^2$Laboratoire Charles Fabry de l'Institut d'Optique -- CNRS -- Univ. Paris-Sud, Campus Polytechnique, RD 128, 91127 Palaiseau Cedex, France}

\begin{abstract}
Continuous-variable quantum key distribution protocols, based on Gaussian modulation of the quadratures of coherent states, have been implemented in recent experiments. A present limitation of such systems is the finite efficiency of the detectors, which can in principle be compensated for by the use of classical optical preamplifiers. Here we study this possibility in detail, by  deriving the modified secret key generation rates when an optical parametric amplifier is placed at the output of the quantum channel. After presenting a general set of security proofs, we show that the use of preamplifiers does compensate for all the imperfections of the detectors when the amplifier is optimal in terms of gain and noise. Imperfect amplifiers can also enhance the system performance, under conditions which are  generally satisfied in practice.
\end{abstract}

\pacs{03.67.Dd, 42.81.-i}

\maketitle
\section{\label{sec:intro}Introduction}

Continuous-variable quantum key distribution (CVQKD) has been proposed in the past few years as a promising alternative to the most commonly implemented discrete-variable QKD. The CVQKD approach has two main advantages: first, it avoids  the limitations associated with single photon counting, and second, it offers the prospect of very high rate secure key distribution. Whereas discrete-variable protocols encode the key information in properties of single photon pulses~\cite{scarani:quantph2008}, CVQKD protocols use for the same purpose optical variables that can take a continuous range of values, such as the quadratures of a mode of the electromagnetic field. The different CVQKD protocols can be roughly categorized by the type of states, modulation, detection system and error-correction algorithm they are employing~\cite{ralph:pra1999,hillery:pra2000,bencheikh:jmo2001,cerf:pra2001,heid:pra2007,zhao:quantph2008}. Here we are interested in protocols where Alice encodes the key information by modulating the quadratures $x$ and $p$ of few-photon coherent states with a centred Gaussian distribution~\cite{grosshans:prl2002,grosshans:nature2003,weedbrook:prl2004}. At the receiver's site, Bob measures either one of the two quadratures randomly using \emph{homodyne detection}~\cite{grosshans:prl2002,grosshans:nature2003}, or both quadratures simultaneously with \emph{heterodyne detection}, thus avoiding the need for random basis switching~\cite{weedbrook:prl2004}. In the final step of the protocol, Alice and Bob apply a reconciliation procedure to form a secret binary key from the continuous information they share.

The security of the Gaussian CVQKD protocol with homodyne detection was first proven against individual Gaussian eavesdropping attacks, using either direct~\cite{grosshans:prl2002} or reverse~\cite{grosshans:nature2003} reconciliation. Security proofs were then obtained against general individual or finite-size attacks~\cite{grosshans:prl2004}, and general collective attacks~\cite{navasques:prl2006,garcia-patron:prl2006,lodewyck:pra2007}. For the protocol with heterodyne detection, the first bounds of Eve's accessible information in the case of individual Gaussian attacks were provided in~\cite{weedbrook:prl2004,weedbrook:pra2006}, and were later improved in~\cite{lodewyck:pra2007b,sudjana:pra2007}, while the case of collective attacks was analyzed in~\cite{garcia-patron:phd2007}. Recently, the unconditional security of both homodyne and heterodyne protocols has also been proven~\cite{renner:quantph2008}.

The theoretical security for systems implementing CVQKD protocols has therefore been proven to be maximal, and is not degraded by the inherent imperfections of Bob's detector, such as noise and inefficiency, that are present in practical CVQKD systems. These imperfections, however, degrade the generation rate of the final secret key. A possible way to overcome this limitation is to use optical parametric amplifiers to boost the signal just before detection, thus compensating at least partially for detector losses~\cite{leonhardt:prl1994}, as has been demonstrated experimentally, for example in \cite{levenson:prl1993,bencheikh:apl1995}. In the context of quantum cryptography \cite{hillery:pra2000,bencheikh:jmo2001}, we therefore need to evaluate in what way the expected improvement in effective quantum efficiency translates into an improvement in secret key generation rate, based on the security proofs mentioned above.

In this paper, we propose to insert an amplifier at the output of the quantum channel and inside Bob's apparatus, and we calculate the resulting secret key generation rate for different cases of eavesdropping attacks. For this purpose, we assume that Bob's apparatus is inaccessible to the eavesdropper, Eve. This assumption is consistent with the idea that if Eve could indeed have access to Bob's setup, then she would also have access to classical data storage, or could impersonate Bob, and the security of the entire communication would be compromised. Since such a situation is unacceptable for a real system, we perform calculations under the ``realistic" assumption that the eavesdropper cannot benefit from Bob's system imperfections.

The paper is organized as follows: In section~\ref{sec:rateswoamp}, we review the expressions of the secret key generation rate for the homodyne protocol, and we derive corresponding expressions for the heterodyne protocol under realistic assumptions, in the case of a standard continuous-variable QKD system with no amplifier. In section~\ref{sec:rateswamp} we add in the system an optical parametric amplifier placed at the output of the quantum channel, and derive the modified secret key generation rates for the homodyne and heterodyne CVQKD protocols, and for both individual and collective attacks. Finally, we compare the performance of practical systems implementing these protocols for different configurations. In general we find that choosing the appropriate combination of detector and amplifier allows us to compensate for all the inherent imperfections of a practical detector if the amplifier has a minimal noise. We also show that a realistic noisy amplifier can also be employed, as long as the noise of the amplifier is small relative to the noise and inefficiency of the detector, which is generally the case in practice.


\section{\label{sec:rateswoamp}Secret key distribution rates for Gaussian CVQKD protocols}

In the following, we first review the basic notions related to continuous-variable QKD protocols with Gaussian modulation and present the assumptions of our calculations. We then derive the expressions of the secret key generation rate for these protocols when homodyne or heterodyne detectors are used, and for the case of individual or collective eavesdropping attacks. The tools that we present here will be essential for the calculations in the next section, which take into account the use of amplifiers to compensate for detector imperfections and thus enhance the performance of CVQKD systems.

\subsection{\label{sec:notations}Notations and assumptions}

The standard prepare-and-measure description of the CVQKD protocol with Gaussian modulation of coherent states was briefly presented in the introduction. In more detail, the quantum transmission phase of the communication between Alice and Bob is described as follows:
\begin{itemize}
\item For each generated pulse, Alice randomly chooses two values for the in-phase quadrature $x_{\subs{A}}$ and the orthogonal quadrature $p_{\subs{A}}$ from a Gaussian distribution centred at zero and of variance $\VA N_0$, where $N_0$ is the shot noise variance. She then prepares a coherent state centred at $(x_{\subs{A}},p_{\subs{A}})$ and sends it to Bob through the quantum channel. The channel features a transmission efficiency $T$ and an excess noise $\eps$, resulting in a noise variance at Bob's input of $(1 + T\eps)N_0$. The total channel-added noise referred to the channel input, expressed in shot noise units, is defined as $\Cl =(1+T\eps)/T -1 = 1/T - 1 + \eps$.
\item When Bob receives the modulated coherent state, he measures either one of the two quadratures randomly (homodyne case) or both quadratures simultaneously (heterodyne case). A practical detector is characterized by an efficiency $\eta$ and a noise $\Velec$ due to detector electronics. As we did for the channel, we can define a detection-added noise referred to Bob's input and expressed in shot-noise units that we denote in general as $\Chi$, and is given by the expressions $\Ch = [(1-\eta) + \Velec]/\eta$ and $\Che = [1 + (1-\eta) + 2\Velec]/\eta$ for homodyne and heterodyne detection, respectively. The total noise referred to the channel input can then be expressed as $\Ct = \Cl + \Chi/T$.
\end{itemize}

\begin{figure*}
  \centering
  \includegraphics[width=0.65\textwidth]{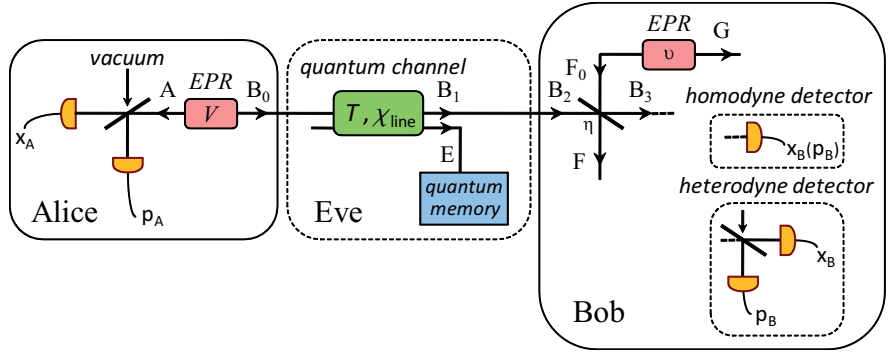}
\caption{Entanglement-based scheme of a Gaussian coherent-state CVQKD protocol with homodyne or heterodyne detection. The transmission $T$ and channel-added noise $\Cl$ are controlled by Eve, who however does not have access to Bob's detection apparatus.}
\label{fig:ebscheme}
\end{figure*}

The prepare-and-measure description presented above is equivalent to the entanglement-based scheme shown in figure \ref{fig:ebscheme}. This equivalence is at the heart of security proofs for this type of CVQKD protocols and has been explained in detail in~\cite{lodewyck:pra2007,garcia-patron:phd2007,grosshans:qic2003}. In this scheme:
\begin{itemize}
\item The coherent state preparation by Alice is modeled by a heterodyne measurement of one half of a two-mode squeezed vacuum (EPR) state of variance $V = \VA + 1$. The other half of the EPR state is sent to Bob through the quantum channel.
\item Bob's detector inefficiency is modeled by a beamsplitter with transmission $\eta$, while its electronic noise $\Velec$ is modeled by an EPR state of variance $v$, one half of which is entering the second input port of the beamsplitter, as shown in figure \ref{fig:ebscheme}. The variance $v$ is chosen to obtain the appropriate expression of $\Chi$, in the following way: for homodyne detection, $v = \eta\Ch/(1-\eta) = 1 + \Velec/(1 - \eta)$, and for heterodyne detection, $v = (\eta\Che - 1)/(1-\eta) = 1 + 2\Velec/(1 - \eta)$, where the 1 in the numerator ($\eta\Che - 1$) of the last expression is subtracted due to the unit of shot noise already introduced by the heterodyne detection.
\end{itemize}

Once the quantum transmission phase of the communication has ended, Alice and Bob proceed with classical data processing procedures, which include a reconciliation algorithm to extract an identical chain of bits from their correlated continuous data, and a standard privacy amplification process to derive a final secret key from this chain. The reconciliation is \emph{direct} when Alice's data is used as a reference for establishing the key and \emph{reverse} when the reference is Bob's data. Reverse reconciliation has been shown to offer a great advantage in QKD system performance~\cite{grosshans:nature2003}, therefore calculations in this paper have been performed for this case. Direct reconciliation expressions can be derived using similar tools as the ones presented here.

Under the assumptions that we have described, we want to calculate the secret key generation rates for the Gaussian coherent-state CVQKD protocol with homodyne and heterodyne detection, for the case of individual and collective eavesdropping attacks. These are considered to be Gaussian attacks, which have been shown to be optimal~\cite{navasques:prl2006,garcia-patron:prl2006}. In the case of \emph{individual attacks}, Eve is authorized to interact individually with each coherent-state pulse sent by Alice, store her ancillae in a quantum memory, and perform measurements on them after sifting (for example, in the case of homodyne detection, after Bob has revealed the quadrature he chose to measure), but before the reconciliation phase. In the case of \emph{collective attacks}, Eve interacts individually with each pulse but is allowed to wait for the entire classical procedure to end before performing the best possible collective measurement on her ensemble of stored ancillae. The maximum information on Bob's key available to Eve is limited by the Shannon bound $\IBE$~\cite{shannon1948,shannon1949} for individual attacks and by the Holevo bound $\xBE$~\cite{holevo1998} for collective attacks. In the following, we will derive expressions for $\IBE$ and $\xBE$ as a function of system parameters. Then, from an information-theoretic perspective, the secret key information that Alice and Bob can distill is defined, in the case of reverse reconciliation, as $\Delta I_{\subs{Shannon}} = \IAB - \IBE$ for individual and $\Delta I_{\subs{Holevo}} = \IAB - \xBE$ for collective attacks, where $\IAB$ is the information shared between Alice and Bob.

It is important to note that the security of the CVQKD protocols against coherent attacks has recently been proven~\cite{renner:quantph2008}. \emph{Coherent attacks} are the most powerful attacks allowed by quantum mechanics. They allow Eve to interact collectively with all the pulses sent by Alice and perform joint measurements on her ancillae after the entire quantum and classical communication. The security proofs show that the derived bounds for the secret key generation rate in the case of collective attacks remain asymptotically valid for arbitrary coherent attacks. Therefore, the results that we derive in the following sections for collective attacks are valid for coherent attacks as well, guaranteeing the unconditional security of the corresponding QKD systems.

\subsection{\label{sec:individual}Individual attacks}

{\bf Homodyne detection case:} The mutual information of Alice and Bob, $\IAB$, is derived from Bob's measured variance $\VB = \eta T(V + \Ct)$ and the conditional variance $V_{\subs{B}|\subs{A}} = \eta T(1 + \Ct)$ using Shannon's equation:
\begin{equation}
\label{eq:IABhom}
\IAB^{\subs{hom}} = \frac{1}{2}\log_2\frac{\VB}{V_{\subs{B}|\subs{A}}} = \frac{1}{2}\log_2\frac{V + \Ct}{1 + \Ct}
\end{equation}
Eve's information on Bob's measured quadrature, $\IBE$, is also derived using Shannon's equation in the case of individual attacks~\cite{grosshans:prl2004}. For reverse reconciliation and under the assumption that Eve cannot benefit from the detection-added noise, $V_{\subs{B}|\subs{E}} = \eta \left[\frac{1}{T(1/V + \Cl)} + \Ch\right]$, and therefore~\cite{lodewyck:pra2007}
\begin{equation}
\label{eq:IBEhom}
\IBE^{\subs{hom}} = \frac{1}{2}\log_2\frac{\VB}{V_{\subs{B}|\subs{E}}} = \frac{T^2(V + \Ct)(1/V + \Cl)}{1 + T\Ch(1/V + \Cl)}
\end{equation}
The Shannon secret key generation rate is then given by $\Delta I_{\subs{Shannon}}^{\subs{hom}} = \IAB^{\subs{hom}} - \IBE^{\subs{hom}}$. \\

{\bf Heterodyne detection case:} It is straightforward to derive the mutual information of Alice and Bob for the case of two measured quadratures:
\begin{equation}
\label{eq:IABhet}
\IAB^{\subs{het}} = 2 \times \frac{1}{2}\log_2\frac{\VB}{V_{\subs{B}|\subs{A}}} = \log_2\frac{V + \Ct}{1 + \Ct}
\end{equation}
where here $\VB = \eta T(V + \Ct)/2$, $V_{\subs{B}|\subs{A}} = \eta T(1 + \Ct)/2$, and $\Ct$ takes the appropriate expression for heterodyne detection (section~\ref{sec:notations}).

The information that Eve gains on Bob's data given that both quadratures are measured is given by $\IBE^{\subs{het}} = \log_2(\VB/V_{\subs{B}|\subs{E}})$. A bound on $V_{\subs{B}|\subs{E}}$ when Eve is allowed to have access to Bob's setup has been calculated in~\cite{lodewyck:pra2007b,sudjana:pra2007}. It is given by the expression $V_{\subs{B}|\subs{E}} = \left(\frac{V x_{\subs{E}} + 1}{V + x_{\subs{E}}} + 1\right)/2$, where $x_{\subs{E}}=T(2-\eps)^2/(\sqrt{2-2T+T\eps}+\sqrt{\eps})^2+1$~\cite{lodewyck:pra2007b}.

To extend this bound to a more realistic scenario, we need to take into account the fact that the detection is not accessible to Eve, and only adds noise that is not correlated to the signal. The corresponding signal-noise commutators are therefore all zero, and the calculation is equivalent to correcting the various variances that intervene in the expressions, to account for the detection parameters. We then find $V_{\subs{B}|\subs{E}} = \eta\left(\frac{V x_{\subs{E}} + 1}{V + x_{\subs{E}}} + \Che\right)/2$, where $x_{\subs{E}}$ is the same as above. Putting things together, Eve's information is in this case given by the following expression:
\begin{equation}
\label{eq:IBEhet}
\IBE^{\subs{het}} = \log_2\frac{\VB}{V_{\subs{B}|\subs{E}}} = \log_2\frac{T(V + \Ct)(V + x_{\subs{E}})}{V x_{\subs{E}} + 1 + \Che(V + x_{\subs{E}})}
\end{equation}
The Shannon secret key generation rate $\Delta I_{\subs{Shannon}}^{\subs{het}}$ is then calculated from (\ref{eq:IABhet}) and (\ref{eq:IBEhet}).

\subsection{\label{sec:collective}Collective attacks}

For convenience, we consider in this section the homodyne and heterodyne cases in parallel. The mutual information between Alice and Bob is given in the case of collective attacks by the same expressions as for individual attacks, namely (\ref{eq:IABhom}) and (\ref{eq:IABhet}) for homodyne and heterodyne detection, respectively. Deriving Eve's information on Bob's measurements, on the other hand, requires a different approach that has been developed in detail in~\cite{lodewyck:pra2007,garcia-patron:phd2007} for the homodyne detection case. Below we present the main ideas of this technique and extend it to the heterodyne detection case.

The maximum information available to Eve on Bob's key is bounded by the Holevo quantity~\cite{renner:phd2005}
\begin{equation}
\label{eq:chiBEHolevo}
\xBE = S(\rho_{\subs{E}})-\int{\rm d}m_{\subs{B}} \; p(m_{\subs{B}}) \; S(\rho_{\subs{E}}^{m_{\subs{B}}})
\end{equation}
where $m_{\subs{B}}$ represents the measurement of Bob, and it can take the form $m_{\subs{B}} = x_{\subs{B}}$ $({\rm d}m_{\subs{B}} = {\rm d}x_{\subs{B}})$ for a homodyne detector or the form $m_{\subs{B}} = x_{\subs{B}}, p_{\subs{B}}$ $({\rm d}m_{\subs{B}} = {\rm d}x_{\subs{B}}{\rm d}p_{\subs{B}})$ for a heterodyne detector. Also, $p(m_{\subs{B}})$ is the probability density of the measurement, $\rho_{\subs{E}}^{m_{\subs{B}}}$ is the eavesdropper's state conditional on Bob's measurement result, and $S$ is the Von Neumann entropy of the quantum state $\rho$.

Using the fact that Eve's system purifies the system AB$_1$, that Bob's measurement purifies the system AEFG (see figure \ref{fig:ebscheme} for mode notation), and that $S(\rho_{\subs{AFG}}^{m_{\subs{B}}})$ is independent of $m_{\subs{B}}$ for Gaussian protocols, $\xBE$ becomes~\cite{lodewyck:pra2007}:
\begin{equation}
\label{eq:chiBEentropies}
\xBE = S(\rho_{\subs{AB}_1})- S(\rho_{\subs{AFG}}^{m_\subs{B}})
\end{equation}
Since it has been shown that Gaussian attacks are optimal for collective attacks~\cite{navasques:prl2006,garcia-patron:prl2006}, it is enough to consider Gaussian states, in which case the expressions for the entropies can be further simplified as follows:
\begin{equation}
\label{eq:chiBEeignevalues}
\xBE = \sum_{i = 1}^2 G\left(\frac{\lambda_i-1}{2}\right) - \sum_{i = 3}^5 G\left(\frac{\lambda_i-1}{2}\right)
\end{equation}
where $G(x)=(x+1)\log_2(x+1)-x\log_2x$, $\lambda_{1,2}$ are the symplectic eigenvalues of the covariance matrix $\gamma_{\subs{AB}_1}$ characterizing the state $\rho_{\subs{AB}_1}$, and $\lambda_{3,4,5}$ are the symplectic eigenvalues of the covariance matrix $\gamma_{\subs{AFG}}^{m_\subs{B}}$ characterizing the state $\rho_{\subs{AFG}}^{m_\subs{B}}$ after Bob's projective measurement.

The covariance matrix $\gamma_{\subs{AB}_1}$ only depends on the system including Alice and the quantum channel, therefore the first part of (\ref{eq:chiBEeignevalues}) is the same for the homodyne and heterodyne cases. The matrix is written as
\begin{eqnarray}
\label{eq:gammaAB1}
 		\gamma_{\subs{AB}_1} & = & \left[\begin{array}{cc}
			\gamma_{\subs{A}} & \sigma_{\subs{AB}_1}^T \\
			\sigma_{\subs{AB}_1} & \gamma_{\subs{B}_1}
			\end{array}
		\right] \\ & = & \left[\begin{array}{cc}
			V \cdot \openone_2 & \sqrt{T(V^2-1)}\cdot \sigma_z \\
			\sqrt{T(V^2-1)}\cdot \sigma_z & T(V+\Cl) \cdot \openone_2
			\end{array}
		\right] \nonumber
\end{eqnarray}
where $\openone_2$ is the $2\times2$ identity matrix and $\sigma_z = \left[\begin{array}{cc}1&0\\0&-1\end{array}\right]$. The symplectic eigenvalues $\lambda_{1,2} \geq 1$ of the above matrix are given by
\begin{eqnarray}
\label{eq:lambda12}
\lambda^{2}_{1,2} & = & \frac{1}{2}\left[A \pm \sqrt{A^2-4B}\right], \mbox{ }\textrm{with} \\
A & = & V^2(1 - 2T) + 2T + T^2(V + \Cl)^2 \nonumber \\
B & = & T^2(V\Cl + 1)^2 \nonumber
\end{eqnarray}

To calculate the second part of (\ref{eq:chiBEeignevalues}), we need to find the symplectic eigenvalues of the covariance matrix $\gamma_{\subs{AFG}}^{m_\subs{B}}$, which can be written as
\begin{equation}
\label{eq:gammaAFGmB}
\gamma_{\subs{AFG}}^{m_\subs{B}} = \gamma_{\subs{AFG}}-\sigma^T_{\subs{AFGB}_3}H\sigma_{\subs{AFGB}_3}
\end{equation}
In the above equation, $H$ is the symplectic matrix that represents the homodyne or heterodyne measurement on mode B$_3$. In the former case, $H_{\subs{hom}} = (X\gamma_{\subs{B}_3}X)^\mathrm{MP}$, where $X = \left[\begin{array}{cc}1&0\\0&0\end{array}\right]$ and MP stands for the Moore-Penrose pseudo-inverse of a matrix, while in the latter case, $H_{\subs{het}} = (\gamma_{\subs{B}_3}+\openone_2)^{-1}$~\cite{garcia-patron:phd2007}. The matrices $\gamma_{\subs{B}_3}$, $\gamma_{\subs{AFG}}$, and $\sigma_{\subs{AFGB}_3}$ can all be derived from the decomposition of the covariance matrix:
\begin{equation}
\label{eq:gammaAFGB3}
 		\gamma_{\subs{AFGB}_3}=\left[\begin{array}{cc}
			\gamma_{\subs{AFG}} & \sigma^T_{\subs{AFGB}_3}\\
			\sigma_{\subs{AFGB}_3} & \gamma_{\subs{B}_3}
			\end{array}
		\right]
\end{equation}
The above matrix can be derived with appropriate rearrangement of lines and columns from the matrix describing the system AB$_3$FG (figure \ref{fig:ebscheme}):
\begin{equation}
\label{eq:gammaAB3FG}
\gamma_{\subs{AB}_3\subs{FG}} = (Y^{\subs{BS}})^T[\gamma_{\subs{AB}_1}\oplus\gamma_{\subs{F}_0\subs{G}}]Y^{\subs{BS}}
\end{equation}
Here, $\gamma_{\subs{AB}_1} (= \gamma_{\subs{AB}_2})$ is given in (\ref{eq:gammaAB1}), while $\gamma_{\subs{F}_0\subs{G}}$ is the matrix that describes the EPR state of variance $v$ used to model the detector's electronic noise. It is written as
\begin{equation}
\label{eq:gammaF0G}
 		\gamma_{\subs{F}_0\subs{G}} = \left[\begin{array}{cc}
			v \cdot \openone_2 & \sqrt{(v^2-1)}\cdot \sigma_z \\
			\sqrt{(v^2-1)}\cdot \sigma_z & v \cdot \openone_2
			\end{array}
		\right]
\end{equation}
where $v$ takes the appropriate value for the homodyne or heterodyne detection case (section~\ref{sec:notations}). Finally, the matrix $Y^{\subs{BS}}$ describes the beamsplitter transformation that models the inefficiency of the detector and acts on modes B$_2$ and F$_0$. It is given by the expression:
\begin{eqnarray}
\label{eq:YBS}
 		Y^{\subs{BS}} & = & \openone_{\subs{A}}\oplus Y_{\subs{B}_2\subs{F}_0}^{\subs{BS}}\oplus\openone_{\subs{G}}, \mbox{ }\textrm{with}\\
        Y_{\subs{B}_2\subs{F}_0}^{\subs{BS}} & = & \left[\begin{array}{cc}
			\sqrt{\eta} \cdot \openone_2 & \sqrt{1 - \eta}\cdot \openone_2 \\
			-\sqrt{1 - \eta}\cdot \openone_2 & \sqrt{\eta} \cdot \openone_2
			\end{array}
		\right] \nonumber
\end{eqnarray}

We now have all the elements required to proceed to the calculation of the symplectic eigenvalues $\lambda_{3,4,5}$. For both homodyne and heterodyne cases, we find that the eigenvalues $\lambda_{3,4} \geq 1$ are given by expressions of the form
\begin{equation}
\label{eq:lambda34}
\lambda^{2}_{3,4}=\frac{1}{2}\left[C \pm \sqrt{C^2-4D}\right]
\end{equation}
where for the homodyne case~\cite{lodewyck:pra2007},
\begin{eqnarray}
\label{eq:CDhom}
C_{\subs{hom}} & = & \frac{A\Ch + V \sqrt{B} + T(V + \Cl)}{T(V + \Ct)} \\
D_{\subs{hom}} & = & \sqrt{B}\frac{V + \sqrt{B}\Ch}{T(V + \Ct)} \nonumber
\end{eqnarray}
and for the heterodyne case,
\begin{eqnarray}
\label{eq:CDhet}
C_{\subs{het}} & = & \frac{1}{(T(V + \Ct))^2}\left[A\Che^2 + B + 1 + \right. \\
& & \left. + 2\Che(V \sqrt{B} + T(V + \Cl)) + 2 T (V^2 - 1)\right] \nonumber \\
D_{\subs{het}} & = & \left(\frac{V + \sqrt{B}\Che}{T(V + \Ct)}\right)^2 \nonumber
\end{eqnarray}
where $A, B$ are given in (\ref{eq:lambda12}). The last symplectic eigenvalue is $\lambda_5 = 1$ for both cases. Based on (\ref{eq:chiBEeignevalues}), (\ref{eq:lambda12}), (\ref{eq:lambda34}), (\ref{eq:CDhom}), and (\ref{eq:CDhet}), we calculate the Holevo information bound $\xBE$ and thus derive the Holevo secret key generation rate $\Delta I_{\subs{Holevo}}^{\subs{hom}} = \IAB^{\subs{hom}} - \xBE^{\subs{hom}}$ and $\Delta I_{\subs{Holevo}}^{\subs{het}} = \IAB^{\subs{het}} - \xBE^{\subs{het}}$ for homodyne and heterodyne detection, respectively.


\section{\label{sec:rateswamp}Adding an amplifier to compensate for detector imperfections}

In the practical case that we are considering, where Bob's detection apparatus has inherent imperfections that degrade the secret key generation rate, it is important to consider ways of overcoming this limitation by compensating for these imperfections. To this end, the use of optical parametric amplifiers, practical and thoroughly studied devices, appears as a natural choice~\cite{hillery:pra2000}. Under the realistic assumption that the amplifier is not available to the eavesdropper, its use can allow Bob to compensate fully or partially for the losses that occur after the output of the amplifier, and thus enhance the system performance in terms of secret key distribution rate and maximal communication distance.

In the following, we first provide models for two types of classical amplifiers and then combine these models with the calculations of section~\ref{sec:rateswoamp} to derive the modified secret key generation rates for Gaussian coherent-state CVQKD protocols with homodyne and heterodyne detection, and for individual and collective eavesdropping attacks, when a classical amplifier is placed at the input of Bob's apparatus. We also apply the results to practical systems and compare their performance for detectors and amplifiers with different characteristics.

\subsection{\label{sec:ampmodel}Amplifier models}

Because of their importance for various technological applications including communication systems, optical amplifiers and their noise characteristics have been studied extensively~\cite{caves:prd1982}. Here we consider two types of amplifiers, an ideal phase-sensitive amplifier and a practical phase-insensitive amplifier. \\

\begin{figure}[b]
  \centering
  \includegraphics[width=0.55\columnwidth]{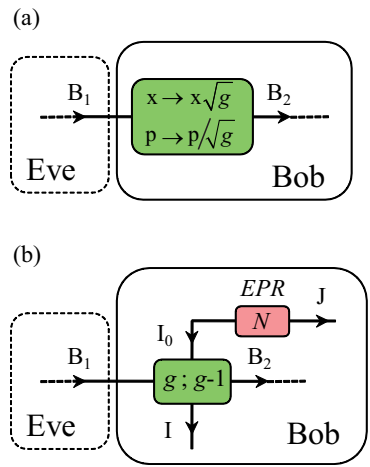}
\caption{Model for (a) an ideal phase-sensitive and (b) a practical phase-insensitive amplifier, placed at the output of the quantum channel and inside Bob's apparatus.}
\label{fig:ampmodels}
\end{figure}

{\bf Phase-sensitive amplifier:} The phase-sensitive amplifier (PSA) is a degenerate optical parametric amplifier that ideally permits noiseless amplification of a chosen quadrature. It is described by the transformations $x_{\subs{S}} \rightarrow \sqrt{g}x_{\subs{S}}$, $p_{\subs{S}} \rightarrow p_{\subs{S}}/\sqrt{g}$, where $g > 1$ is the gain of the amplification and $x_{\subs{S}}, p_{\subs{S}}$ are the two quadratures of the signal mode. Therefore, the amplification process has an asymmetric effect on the quadratures, with the in-phase one being amplified and the orthogonal one squeezed. A realistic amplifier can also add noise to this process but this effect is neglected here.

Modeling this type of amplifier for our purposes is straightforward. The model is shown in figure \ref{fig:ampmodels}(a), where the mode notation refers to figure \ref{fig:ebscheme}. \\

{\bf Phase-insensitive amplifier:} The phase-insensitive amplifier (PIA) is a non-degenerate optical parametric amplifier, which is described by the transformations $x_{\subs{S}} \rightarrow \sqrt{g}x_{\subs{S}} + \sqrt{g - 1}x_{\subs{I}}$, $p_{\subs{S}} \rightarrow \sqrt{g}p_{\subs{S}} - \sqrt{g - 1}p_{\subs{I}}$, and $x_{\subs{I}} \rightarrow \sqrt{g}x_{\subs{I}} + \sqrt{g - 1}x_{\subs{S}}$, $p_{\subs{I}} \rightarrow \sqrt{g}p_{\subs{I}} - \sqrt{g - 1}p_{\subs{S}}$. Here, $g > 1$ is again the gain of amplification, while S and I denote the signal mode and an idler mode that is ideally in a vacuum state or, in a more realistic setting, in a state featuring a noise variance $V_{\subs{I}} = N > 1$ (in shot-noise units). Therefore, this type of amplifier amplifies symmetrically both quadratures but the amplification process is associated with a fundamental excess noise that arises from the coupling of the signal input to the internal modes of the amplifier~\cite{caves:prd1982,ou:prl1993}. When these are in a vacuum state, $N = 1$ and the phase-insensitive amplifier adds a minimal noise for a given gain $g$. Such an amplifier can be approximated in practice using nonlinear processes in optical crystals~\cite{ou:prl1993,levenson:josab1993} or doped optical fibres~\cite{fasel:prl2002}.

Figure \ref{fig:ampmodels}(b) shows the model for the phase-insensitive amplifier. It consists of a noiseless amplifier that applies the appropriate gain factor to each input mode and an EPR state of variance $N$, one half of which is entering the amplifier's second input port, and which serves the purpose of modeling the amplifier's inherent noise.

\subsection{\label{sec:modrates}Modified secret key generation rates}

Considering the characteristics of the homodyne and heterodyne detectors on the one hand and of the phase-sensitive and phase-insensitive amplifiers on the other, it is natural to expect that certain configurations are better adapted than others. Indeed, the symmetrical amplifying effect of a phase-insensitive amplifier on the two quadratures of its input mode appears to be more suitable for a heterodyne detector where both quadratures are measured, while the single measured quadrature of a homodyne detector can benefit from the ideally noiseless amplification induced by a phase-sensitive amplifier. We expect then that these two configurations can result in a significantly improved performance of the corresponding CVQKD systems. Consequently, we provide an analysis for these two cases in this section, leaving the other two possible combinations for the appendix.\\

{\bf Homodyne detection and phase-sensitive amplifier case:} We start our analysis with the case of collective attacks when the ideal phase-sensitive amplifier described before is inserted into a CVQKD system with homodyne detection.

As we mentioned in section~\ref{sec:collective}, the information that Eve has gained on Bob's key is given by (\ref{eq:chiBEeignevalues}). The first part of this equation does not depend on Bob's setup, and therefore the results of (\ref{eq:lambda12}) remain unchanged for all cases. For the second part of (\ref{eq:chiBEeignevalues}), we need to calculate the symplectic eigenvalues of the covariance matrix $\gamma_{\subs{AFG}}^{x_\subs{B}}$. This calculation leads to the matrix $\gamma_{\subs{AB}_3\subs{FG}}$ of (\ref{eq:gammaAB3FG}), which is now written as follows:
\begin{equation}
\label{eq:gammaAB3FGmod}
\gamma_{\subs{AB}_3\subs{FG}} = (Y^{\subs{BS}})^T(Y^{\subs{PSA}})^T[\gamma_{\subs{AB}_1}\oplus\gamma_{\subs{F}_0\subs{G}}]Y^{\subs{PSA}}Y^{\subs{BS}}
\end{equation}
In the above equation, the matrices $Y^{\subs{BS}}, \gamma_{\subs{AB}_1}, \gamma_{\subs{F}_0\subs{G}}$ are given in section~\ref{sec:collective}, while the matrix $Y^{\subs{PSA}}$ describes the transformation induced on mode B$_1$ by the phase-sensitive amplifier:
\begin{eqnarray}
\label{eq:YPSA}
 		Y^{\subs{PSA}} & = & \openone_{\subs{A}}\oplus Y_{\subs{B}_1}^{\subs{PSA}}\oplus\openone_{\subs{F}_0}\oplus\openone_{\subs{G}}, \mbox{ }\textrm{with}\\
        Y_{\subs{B}_1}^{\subs{PSA}} & = & \left[\begin{array}{cc}
			\sqrt{g} & 0 \\
			0 & 1/\sqrt{g}
			\end{array}
		\right] \nonumber
\end{eqnarray}
It is now straightforward to derive the modified symplectic eigenvalues. We find that these are actually given again by the expressions of (\ref{eq:lambda34}), (\ref{eq:CDhom}), with the only difference that the detection-added noise $\Ch$ (and consequently the total noise $\Ct$) is conveniently modified to include the effect of the amplifier:
\begin{equation}
\label{eq:xhomPSA}
\Ch^{\subs{PSA}} = \frac{(1 - \eta) + \Velec}{g\eta}
\end{equation}
It is important to note that we have implicitly assumed that Bob always amplifies the quadrature that he has randomly chosen to measure: this is clearly advantageous for him, and in theory straightforward to achieve. However, this requires in practice to phase lock the optical beam pumping the amplifier and the local oscillator of the homodyne detection, which may not be easy to implement.

Concerning the mutual information between Eve and Bob for the case of individual attacks, as well as the information shared between Alice and Bob, we find that the expressions of (\ref{eq:IABhom}) and (\ref{eq:IBEhom}) remain valid with $\Ch$ and $\Ct$ modified as in (\ref{eq:xhomPSA}). \\

{\bf Heterodyne detection and phase-insensitive amplifier case:} As with the previous case, we follow the analysis of section~\ref{sec:collective} for collective attacks, but for heterodyne detection and with a phase-insensitive amplifier placed at the output of the quantum channel.

It is clear from figure \ref{fig:ampmodels}(b) that in order to take into account the amplifier-added noise in this case, two additional modes I and J need to be included in the calculation. Then, $\xBE$ is calculated from the following equations that replace (\ref{eq:chiBEentropies}) and (\ref{eq:chiBEeignevalues}):
\begin{eqnarray}
\label{eq:chiBEmod}
\xBE & = & S(\rho_{\subs{AB}_1})- S(\rho_{\subs{AIJFG}}^{x_\subs{B},p_\subs{B}}) \\
\xBE & = & \sum_{i = 1}^2 G\left(\frac{\lambda_i-1}{2}\right) - \sum_{i = 3}^7 G\left(\frac{\lambda_i-1}{2}\right)
\end{eqnarray}
It is therefore necessary to derive the symplectic eigenvalues of the 5-mode covariance matrix $\gamma_{\subs{AIJFG}}^{x_\subs{B},p_\subs{B}}$, which involves the solution of a polynomial of degree 5. Proceeding as in section~\ref{sec:collective}, we write
\begin{equation}
\label{eq:gammaAIJFGxBpB}
\gamma_{\subs{AIJFG}}^{x_\subs{B},p_\subs{B}} = \gamma_{\subs{AIJFG}}-\sigma^T_{\subs{AIJFGB}_3}H_{\subs{het}}\sigma_{\subs{AIJFGB}_3}
\end{equation}
where $H_{\subs{het}} = (\gamma_{\subs{B}_3}+\openone_2)^{-1}$. The identity matrix in this expression represents the beamsplitter included in the heterodyne detector. As in (\ref{eq:gammaAFGB3}), the matrices in this equation can be derived from the decomposition of the matrix $\gamma_{\subs{AIJFGB}_3}$, which can in its turn be derived by first calculating the matrix (see figures~\ref{fig:ebscheme} and \ref{fig:ampmodels}(b) for mode notation)
\begin{equation}
\label{eq:gammaAB2IJF0G}
\gamma_{\subs{AB}_2\subs{IJF}_0\subs{G}} = (Y^{\subs{PIA}})^T[\gamma_{\subs{AB}_1}\oplus\gamma_{\subs{I}_0\subs{J}}\oplus\gamma_{\subs{F}_0\subs{G}}]Y^{\subs{PIA}}
\end{equation}
then rearranging it to obtain $\gamma_{\subs{AB}_2\subs{F}_0\subs{IJG}}$, and finally calculating
\begin{eqnarray}
\label{eq:gammaAB3FIJG}
\gamma_{\subs{AB}_3\subs{FIJG}} & = &  (Y^{\subs{BS}})^T\gamma_{\subs{AB}_2\subs{F}_0\subs{IJG}} Y^{\subs{BS}}
\end{eqnarray}
to obtain the desired result by a last rearrangement of matrix lines and columns. In the above equations, $\gamma_{\subs{AB}_1}$ and $\gamma_{\subs{F}_0\subs{G}}$ are known from (\ref{eq:gammaAB1}) and (\ref{eq:gammaF0G}), while the beamsplitter transformation is now written as $Y^{\subs{BS}} = \openone_{\subs{A}}\oplus Y_{\subs{B}_2\subs{F}_0}^{\subs{BS}}\oplus\openone_{\subs{I}}\oplus\openone_{\subs{J}}\oplus\openone_{\subs{G}}$ to account for the new modes. Furthermore, the matrix $\gamma_{\subs{I}_0\subs{J}}$ describes the EPR state of variance $N$ used to model the amplifier's inherent noise, and is written as
\begin{equation}
\label{eq:gammaI0J}
 		\gamma_{\subs{I}_0\subs{J}} = \left[\begin{array}{cc}
			N \cdot \openone_2 & \sqrt{N^2-1}\cdot \sigma_z \\
			\sqrt{N^2-1}\cdot \sigma_z & N \cdot \openone_2
			\end{array}
		\right]
\end{equation}
Finally, the transformation induced by the phase-insensitive amplifier on modes B$_1$ and I$_0$ is described by the matrix:
\begin{eqnarray}
\label{eq:YPIA}
 		Y^{\subs{PIA}} & = & \openone_{\subs{A}}\oplus Y_{\subs{B}_1\subs{I}_0}^{\subs{PIA}}\oplus\openone_{\subs{J}}\oplus\openone_{\subs{F}_0}\oplus\openone_{\subs{G}}, \mbox{ }\textrm{with}\\
        Y_{\subs{B}_1\subs{I}_0}^{\subs{PIA}} & = & \left[\begin{array}{cc}
			\sqrt{g} \cdot \openone_2 & \sqrt{g - 1}\cdot \sigma_z \\
			\sqrt{g - 1}\cdot \sigma_z & \sqrt{g} \cdot \openone_2
			\end{array}
		\right] \nonumber
\end{eqnarray}
Based on the above additional elements, a rather complicated but straightforward calculation leads to the expressions of (\ref{eq:lambda34}), (\ref{eq:CDhet}) for the symplectic eigenvalues $\lambda_{3,4}$, while $\lambda_{5,6,7} = 1$. Similarly to the case of homodyne detection with a phase-sensitive amplifier, the effect of the phase-insensitive amplifier can be incorporated into a modified heterodyne detection-added noise:
\begin{equation}
\label{eq:xhetPIA}
\Che^{\subs{PIA}} = \frac{1 + (1-\eta) + 2\Velec + N(g-1)\eta}{g\eta}
\end{equation}

For individual attacks, as in the previous case, the Shannon expressions of (\ref{eq:IABhet}) and (\ref{eq:IBEhet}) remain the same with $\Che$ and $\Ct$ modified as in (\ref{eq:xhetPIA}).

\subsection{\label{sec:application}Application to practical systems}

In this section, we apply the results derived in section~\ref{sec:modrates} to practical QKD systems in order to compare their performance for different configurations. In particular, we calculate the secret key generation rate as a function of distance for fibre-optic implementations of the Gaussian coherent-state CVQKD protocol,  for individual or collective eavesdropping attacks, in two possible configurations: homodyne detection with a phase-sensitive amplifier placed at the output of the quantum channel, and heterodyne detection with a phase-insensitive amplifier  included in the system. The remaining two detector-amplifier configurations are discussed in the appendix.

The parameters that intervene in the equations that we derived in the previous sections are the variance of Alice's modulation $\VA$, the transmission efficiency $T$ and excess noise $\eps$ of the quantum channel, the efficiency $\eta$ and electronic noise $\Velec$ of the detector, the gain $g$ and potentially the noise $N$ of the amplifier. The parameters $\eps$, $\eta$, and $\Velec$ are fixed in all simulations to the values $\eps = 0.005$ (in shot-noise units), $\eta = 0.6$, and $\Velec = 0.05$ (in shot-noise units), which are standard in CVQKD experiments~\cite{lodewyck:pra2007}. The gain of the amplifier $g$ takes the values 1 (equivalent to no amplifier), 3, or 20, while we set the noise of the phase-insensitive amplifier $N$ to either 1 for minimal (vacuum) noise, or to the more realistic value 1.5 (in shot-noise units, referred to the input). Furthermore, the channel transmission efficiency is written as $T = 10^{-\alpha L/10}$, where $\alpha = 0.2$~dB/km is the loss coefficient of optical fibres, and $L$ is the length of the channel, as a function of which we calculate the secret key generation rate. Finally, the modulation variance $\VA$ is linked to the signal-to-noise ratio (SNR) via Shannon's equation as follows:
\begin{eqnarray}
\label{eq:IABSNR}
\IAB^{\subs{hom}} & = & \frac{1}{2}\log_2(1 + \textrm{SNR}) = \frac{1}{2}\log_2\frac{V + \Ct}{1 + \Ct} \\
& = & \frac{1}{2}\log_2\left(1 + \frac{\VA}{1 + \Ct}\right)\rightarrow \VA = \textrm{SNR} (1 + \Ct) \nonumber
\end{eqnarray}

In the following simulations, we use the SNR as an adjustable parameter, with respect to which we numerically optimize the secret key generation rate for each distance. In this way, we find for each channel length the optimal modulation variance that maximizes the rate given specific system parameters. This optimization is easy to achieve in practical systems because it is in general straightforward to dynamically adjust $\VA$. The SNR range used for the optimization is [0.5;15], which corresponds to values that can be easily reached in practical systems.

\begin{figure}
  \centering
  \includegraphics[width=\columnwidth]{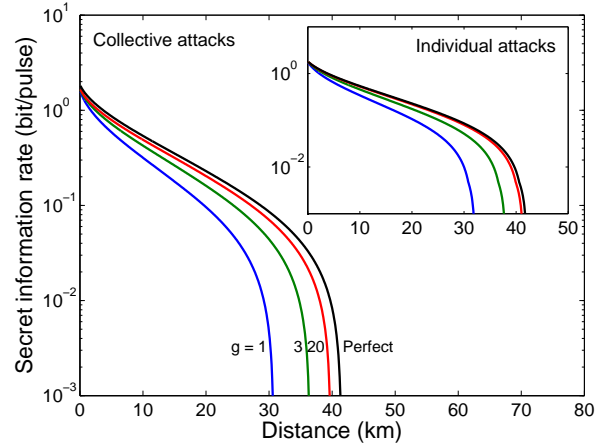}
\caption{Secret key generation rate as a function of distance for a protocol with \emph{homodyne detection} and a \emph{phase-sensitive amplifier}, in the case of collective (main figure) and individual (inset) eavesdropping attacks. The `perfect' curve corresponds to a perfect homodyne detector ($\eta = 1$, $\Velec = 0$) and no amplifier.}
\label{fig:hompsashho}
\end{figure}

\begin{figure}
  \centering
  \includegraphics[width=\columnwidth]{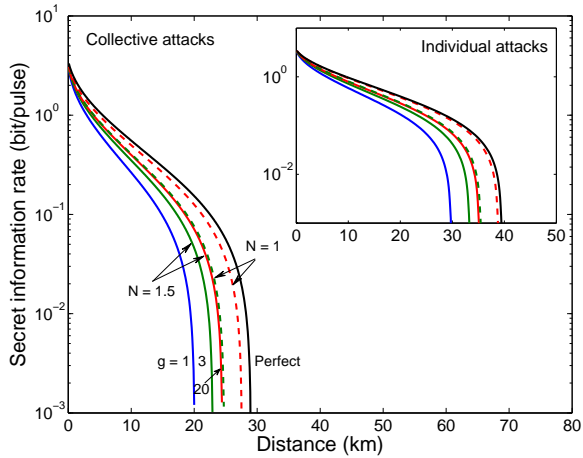}
\caption{Secret key generation rate as a function of distance for a protocol with \emph{heterodyne detection} and a \emph{phase-insensitive amplifier}, for collective and individual attacks. The `perfect' curve corresponds again to a perfect heterodyne detector ($\eta = 1$, $\Velec = 0$) and no amplifier. The amplifier can be either optimal ($N = 1$, dashed lines), or realistic ($N = 1.5$, full lines).}
\label{fig:hetpiashho}
\end{figure}

In addition to the parameters mentioned above, we also take into account the fact that the reconciliation phase of the QKD protocol has in practical systems a finite efficiency. This effect is typically included using a reconciliation efficiency parameter $\beta$ that degrades in all cases the secret key generation rate as $\Delta I_{\subs{Shannon}} = \beta\IAB - \IBE$, or $\Delta I_{\subs{Holevo}}= \beta\IAB - \xBE$. Our empirical work has shown that the maximal practically attainable $\beta$ depends on the SNR (see figure 4 in \cite{lodewyck:pra2007}). For optimization purposes, we have approximated this dependence with the following analytical function:
\begin{equation}
\label{eq:beta}
\beta = \frac{\log(1 + \textrm{SNR}^{1.2})}{1.29\log(1 + \textrm{SNR})} + 0.02
\end{equation}
This function reflects the fact that high reconciliation efficiency can be achieved when working at high SNR, while operating the system at lower SNR values results in a lower reconciliation efficiency. In the SNR range [0.5;15] that we are using, $\beta$ is an increasing function of SNR, while the maximal available secret key rates $I_\subs{AB} - I_\subs{BE}$ or $I_\subs{AB} - \chi_\subs{BE}$ are decreasing functions of SNR~\cite{leverrier:pra2008}. In order to extract the optimal secret key generation rate at a given distance, it is therefore necessary to realize a non-trivial optimization with respect to  the SNR.

The simulation results are shown in figures~\ref{fig:hompsashho} and \ref{fig:hetpiashho} for the two cases presented in section~\ref{sec:modrates}, respectively. We observe that in both cases the effect of the amplifier on the system performance may have a different scale but is essentially the same for collective and individual attacks. Furthermore, the larger the amplification gain the more pronounced this effect is, while in the case of a phase-insensitive amplifier an additional noise naturally degrades the secret key generation rate.

A close examination of the modified detection-added noise expressions derived in section~\ref{sec:modrates} can facilitate the interpretation of the results. More specifically, starting from (\ref{eq:xhomPSA}) for the case of homodyne detection with a phase-sensitive amplifier, we observe that in the limit of large amplification gain this expression tends to zero. Therefore, an ideal phase-sensitive amplifier can compensate for all the imperfections of a practical homodyne detector, that is from a system perspective their combination is equivalent to a perfect detection apparatus. Indeed, figure \ref{fig:hompsashho} shows that for large $g$ the secret key generation rate approaches that of the corresponding perfect detector curve.

For heterodyne detection with a phase-insensitive amplifier, the expression of (\ref{eq:xhetPIA}) that applies to this case tends to $N$ in the limit of large gain. We can see then that the combined amplifier and practical detector can be equivalently described by a noiseless heterodyne detector featuring a limited efficiency $\eta'$, such that $[1 + (1-\eta')]/\eta' = N$. As we discussed in section~\ref{sec:ampmodel}, for an ideal phase-insensitive amplifier, $N = 1$, which leads to an equivalent detector efficiency $\eta' = 1$. Therefore, an ideal phase-insensitive amplifier can compensate for all the imperfections of a practical heterodyne detector, exactly in the same way that an ideal phase-sensitive amplifier does for a homodyne detector. In a sense, the phase-insensitive amplifier precompensates for the inherent loss due to the beamsplitter of the heterodyne detector. Their combination simulates then a perfect detection apparatus, as shown in figure \ref{fig:hetpiashho}. Clearly, the effect of a more realistic amplifier with $N > 1$ on the system performance depends on the value of this noise. For the specific detector parameters that we are considering, this effect becomes negative when the noise exceeds $2.5 N_0$.

\begin{figure}
  \centering
  \includegraphics[width=\columnwidth]{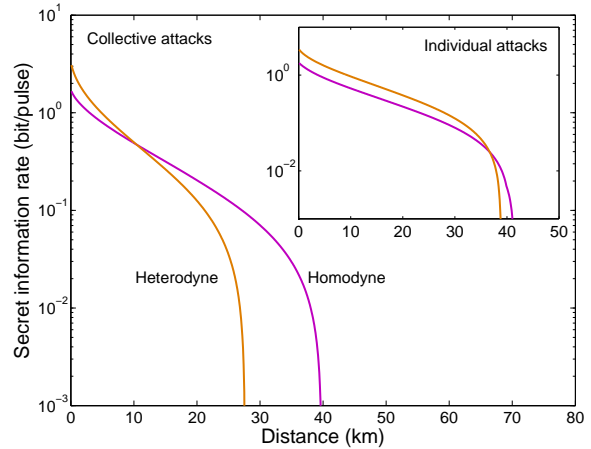}
\caption{Secret key generation rate as a function of distance for protocols with \emph{homodyne detection} and an ideal \emph{phase-sensitive amplifier} with gain $g = 20$, and \emph{heterodyne detection} and an ideal \emph{phase-insensitive amplifier} with gain $g = 20$ and noise $N = 1$, in the case of collective (main figure) and individual (inset) eavesdropping attacks.}
\label{fig:homhetshho}
\end{figure}

From a practical point of view, it is also interesting to directly compare the performance of the homodyne and heterodyne protocols in the configurations previously discussed. In figure \ref{fig:homhetshho} we perform this comparison for ideal amplifiers with gain $g = 20$. We observe that, despite the improvement due to the presence of the amplifier, the heterodyne protocol is in general more vulnerable to system imperfections than the protocol with homodyne detection, under the realistic assumptions we consider in this work. The superior performance of the homodyne protocol is especially pronounced in the case of collective attacks.


\section{\label{sec:conclusion}Conclusion}

In this paper, we have studied continuous-variable quantum key distribution protocols that use Gaussian modulation of coherent states for key encoding and employ homodyne or heterodyne detection techniques. In particular, based on an equivalent entanglement-based scheme for this type of protocols, we reviewed the security of the homodyne protocol against individual and collective eavesdropping attacks, and provided a security analysis for the heterodyne protocol. In all cases, we assumed that the eavesdropper does not have access to Bob's setup, and calculated the secret key generation rates for reverse reconciliation. Subsequently, we studied the effect of adding a classical optical preamplifier at the output of the quantum channel and inside Bob's apparatus to the performance of the QKD system. To this end, we considered two types of amplifiers, an ideal phase-sensitive amplifier and a practical phase-insensitive amplifier, and derived the modified expressions for the secret key generation rate for the various configurations. We then applied these expressions to practical systems and compared the performance of such systems under realistic conditions.

We find that an optimal phase-sensitive amplifier can compensate for all imperfections, such as losses and noise, of a practical homodyne detector, and an optimal phase-insensitive amplifier can do the same for a heterodyne detector, thus enhancing the performance of the corresponding CVQKD systems. Furthermore, our results show that realistic noisy amplifiers can also be employed in such systems with positive effects, as long as their noise does not exceed a certain value that depends on the detector parameters. For practical detectors and amplifiers, this condition is in general satisfied. It is therefore clear that overcoming the limitation imposed by the imperfections of realistic detectors on the secret key rate of CVQKD systems is indeed possible using practical and well studied devices such as optical parametric amplifiers.


\begin{acknowledgements}
The authors thank Anthony Leverrier for helpful comments and suggestions. Financial support was provided by the Integrated European Project SECOQC (Grant No. IST-2002-506813) and the French National Research Agency Project SEQURE. E.D. acknowledges support from the European Union through a Marie-Curie fellowship (MEIF-2006-039719) and a Marie-Curie reintegration grant (MIRG-2006-041265).
\end{acknowledgements}

\appendix
\section*{Appendix}
\setcounter{section}{1}

We provide here the modified secret key generation rates for the detector-amplifier configurations that were not discussed in section~\ref{sec:modrates}, namely the cases of homodyne detection with a phase-insensitive amplifier and heterodyne detection with a phase-sensitive amplifier. We also apply the results for these cases to practical systems and discuss the performance of the corresponding implementations.\\

{\bf Homodyne detection and phase-insensitive amplifier case:} The analysis for collective attacks in this case is essentially the same as in the case of heterodyne detection with this type of amplifier, except for the expression for $\gamma_{\subs{F}_0\subs{G}}$ of (\ref{eq:gammaF0G}), which needs to include the appropriate value for $v$, and the expression of (\ref{eq:gammaAIJFGxBpB}), which is now written as
\begin{equation}
\label{eq:gammaAIJFGxB}
\gamma_{\subs{AIJFG}}^{x_\subs{B}} = \gamma_{\subs{AIJFG}}-\sigma^T_{\subs{AIJFGB}_3}H_{\subs{hom}}\sigma_{\subs{AIJFGB}_3}
\end{equation}
In the above expression, $H_{\subs{hom}} = (X\gamma_{\subs{B}_3}X)^\mathrm{MP}$. Taking into account these changes, we derive the symplectic eigenvalues  $\lambda_{3,4}$ given in (\ref{eq:lambda34}), (\ref{eq:CDhom}), and $\lambda_{5,6,7} = 1$. In this case as well, the effect of the phase-insensitive amplifier can be included in a modified detection-added noise:
\begin{equation}
\label{eq:xhomPIA}
\Ch^{\subs{PIA}} = \frac{(1-\eta) + \Velec + N(g-1)\eta }{g\eta}
\end{equation}

Similarly to the cases presented in section~\ref{sec:modrates}, the mutual information of (\ref{eq:IABhom}) and the Shannon information bound of (\ref{eq:IBEhom}) remain the same, with the appropriate modification of $\Ch$ and $\Ct$. \\

\begin{figure}[b]
  \centering
  \includegraphics[width=\columnwidth]{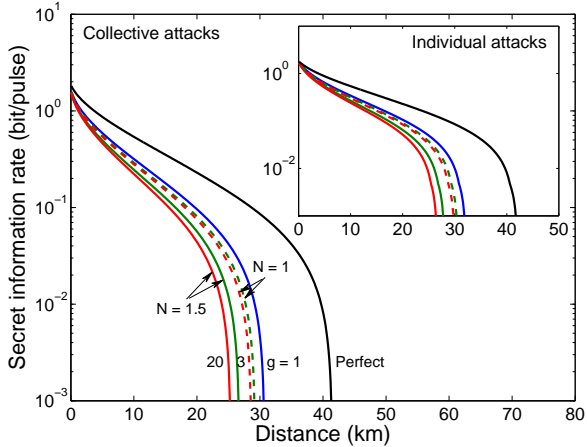}
\caption{Secret key generation rate as a function of distance for a protocol with \emph{homodyne detection} and a \emph{phase-insensitive amplifier}, for collective and individual attacks. The `perfect' curve corresponds again to a perfect homodyne detector ($\eta = 1$, $\Velec = 0$) and no amplifier. The amplifier can be either optimal ($N = 1$, dashed lines), or realistic ($N = 1.5$, full lines).}
\label{fig:hompiashho}
\end{figure}

{\bf Heterodyne detection and phase-sensitive amplifier case:} For a protocol with heterodyne detection when a phase-sensitive amplifier is added in the system, the analysis for collective attacks leads to the expressions of (\ref{eq:lambda34}), (\ref{eq:CDhet}) modified in one important way related to the fact that the amplifier's effect on the two quadratures is different. Since both quadratures are measured, this fact is taken into account by separately defining the modified detection-added noise for quadratures $x$ and $p$ as follows:
\begin{eqnarray}
\label{eq:xhetPSA}
\Che^{\subs{PSA},x} & = & \frac{1 + (1-\eta) + 2 \Velec}{g\eta} \\
\Che^{\subs{PSA},p} & = & g\frac{1 + (1-\eta) + 2 \Velec}{\eta} \nonumber
\end{eqnarray}
Then, the total noise is also correspondingly defined as $\Ct^{x,p} = \Cl + \Che^{\subs{PSA},x,p}/T$, and (\ref{eq:CDhet}) becomes
\begin{eqnarray}
\label{eq:CDhetmod}
C_{\subs{het}}^{\subs{PSA}} & = & \frac{1}{\left(T\left(V + \Ct^x\right)\right)\left(T\left(V + \Ct^p\right)\right)} \times \\
& & \left[A\Che^{\subs{PSA},x}\Che^{\subs{PSA},p} + B + 1 + \right. \nonumber \\
& & + \left(\Che^{\subs{PSA},x} + \Che^{\subs{PSA},p}\right)\left(V \sqrt{B} + T\left(V + \Cl\right)\right) + \nonumber \\
& & \left. + 2 T \left(V^2 - 1\right)\right] \nonumber \\
D_{\subs{het}}^{\subs{PSA}} & = & \left(\frac{V + \sqrt{B}\Che^{\subs{PSA},x}}{T(V + \Ct^x)}\right)\left(\frac{V + \sqrt{B}\Che^{\subs{PSA},p}}{T(V + \Ct^p)}\right) \nonumber
\end{eqnarray}

\begin{figure}[t]
  \centering
  \includegraphics[width=\columnwidth]{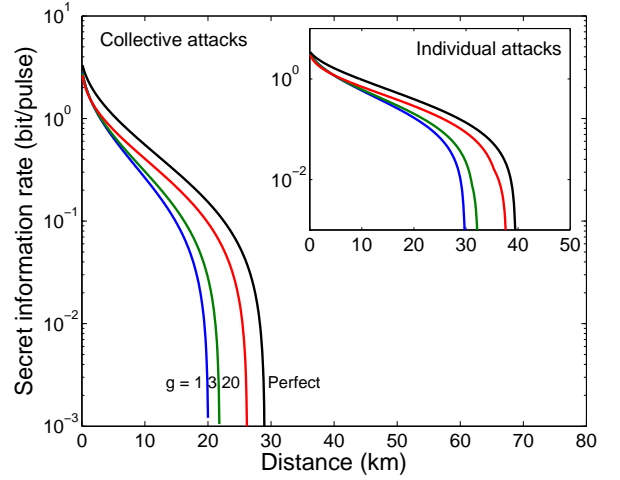}
\caption{Secret key generation rate as a function of distance for a protocol with \emph{heterodyne detection} and a \emph{phase-sensitive amplifier}, in the case of collective and individual attacks. The `perfect' curve corresponds to a perfect heterodyne detector ($\eta = 1$, $\Velec = 0$) and no amplifier.}
\label{fig:hetpsashho}
\end{figure}

The mutual information and Shannon information bound of (\ref{eq:IABhet}) and (\ref{eq:IBEhet}) are modified in a similar way. In particular, we now have
\begin{eqnarray}
\label{eq:VBVBAVBE}
\VB & = & \frac{\eta T}{2}\left[(V + \Ct^x)(V + \Ct^p)\right]^{\frac{1}{2}} \\
V_{\subs{B}|\subs{A}} & = & \frac{\eta T}{2}\left[(1 + \Ct^x)(1 + \Ct^p)\right]^{\frac{1}{2}} \nonumber \\
V_{\subs{B}|\subs{E}} & = & \frac{\eta}{2}\left[\left(\frac{V x_{\subs{E}} + 1}{V + x_{\subs{E}}} + \Che^{\subs{PSA},x}\right)\left(\frac{V x_{\subs{E}} + 1}{V + x_{\subs{E}}} + \Che^{\subs{PSA},p}\right)\right]^{\frac{1}{2}} \nonumber
\end{eqnarray}
where $x_{\subs{E}}$ is given in section~\ref{sec:individual}, and for example the last expression has been calculated from $V_{\subs{B}|\subs{E}} = (V_{\subs{B}|\subs{E}}^x V_{\subs{B}|\subs{E}}^p)^{1/2}$ with
\begin{eqnarray}
\label{eq:VBExVBEp}
V_{\subs{B}|\subs{E}}^x & = & \frac{g\eta}{2}\left(\frac{V x_{\subs{E}} + 1}{V + x_{\subs{E}}} + \Che^{\subs{PSA},x}\right) \\
V_{\subs{B}|\subs{E}}^p & = & \frac{\eta}{2g}\left(\frac{V x_{\subs{E}} + 1}{V + x_{\subs{E}}} + \Che^{\subs{PSA},p}\right) \nonumber
\end{eqnarray}
We then use the standard Shannon equations to derive $\IAB^{\subs{het}}$ and $\IBE^{\subs{het}}$.\\

{\bf Application to practical systems:} Figures~\ref{fig:hompiashho} and \ref{fig:hetpsashho} show the simulation results for the above cases, under the same conditions as the ones detailed in section~\ref{sec:application}. Similarly to the case of heterodyne detection with a phase-insensitive amplifier, the expression of (\ref{eq:xhomPIA}) for homodyne detection with this type of amplifier tends to $N$ in the limit of large gain. In this case, however, the efficiency of the equivalent noiseless homodyne detector satisfies the relationship $(1-\eta')/\eta' = N$, hence $N = 1$ leads to $\eta' = 0.5$. Therefore, even for this optimal scenario, introducing such an amplifier actually degrades the QKD system performance, unless the homodyne detector is untypically noisy and lossy. This negative effect is illustrated in figure \ref{fig:hompiashho}. The case of heterodyne detection with a phase-sensitive amplifier is more subtle, because of the asymmetry expressed in (\ref{eq:xhetPSA}). The effect of introducing an amplifier in this case entirely depends on the system parameters, and is positive for the parameters of figure \ref{fig:hetpsashho}.



\end{document}